\documentclass[aps,prl,twocolumn,superscriptaddress,showpacs,showkeys]{revtex4-2}

\usepackage{color}
\usepackage{graphicx}
\usepackage{dcolumn}
\usepackage{bm}
\usepackage{float}
\usepackage[mathlines]{lineno}
\usepackage{tabularx}
\usepackage[colorlinks,linkcolor=blue,anchorcolor=blue,citecolor=blue,urlcolor=blue,hyperindex,CJKbookmarks]{hyperref}
\usepackage{footnote}
\usepackage{amsmath}
\usepackage{txfonts}
\usepackage{mathptmx}
\usepackage{ragged2e}
\usepackage{booktabs,makecell, multirow, tabularx}
\usepackage{multirow}
\usepackage{braket}
\usepackage{gensymb}
\usepackage{array}
\usepackage{physics} 
\usepackage{textcomp}
\usepackage[T1]{fontenc}

\begin{document}

\title{Strongly correlated altermagnet CaCrO$_3$}

\author{Zhenfeng Ouyang}\affiliation{School of Physics and Beijing Key Laboratory of Opto-electronic Functional Materials $\&$ Micro-nano Devices, Renmin University of China, Beijing 100872, China}\affiliation{Key Laboratory of Quantum State Construction and Manipulation (Ministry of Education), Renmin University of China, Beijing 100872, China}

\author{Peng-Jie Guo}\email{guopengjie@ruc.edu.cn}\affiliation{School of Physics and Beijing Key Laboratory of Opto-electronic Functional Materials $\&$ Micro-nano Devices, Renmin University of China, Beijing 100872, China}\affiliation{Key Laboratory of Quantum State Construction and Manipulation (Ministry of Education), Renmin University of China, Beijing 100872, China}

\author{Rong-Qiang He}\email{rqhe@ruc.edu.cn}\affiliation{School of Physics and Beijing Key Laboratory of Opto-electronic Functional Materials $\&$ Micro-nano Devices, Renmin University of China, Beijing 100872, China}\affiliation{Key Laboratory of Quantum State Construction and Manipulation (Ministry of Education), Renmin University of China, Beijing 100872, China}

\author{Zhong-Yi Lu}\email{zlu@ruc.edu.cn}\affiliation{School of Physics and Beijing Key Laboratory of Opto-electronic Functional Materials $\&$ Micro-nano Devices, Renmin University of China, Beijing 100872, China}\affiliation{Key Laboratory of Quantum State Construction and Manipulation (Ministry of Education), Renmin University of China, Beijing 100872, China}\affiliation{Hefei National Laboratory, Hefei 230088, China}

\date{\today}

\begin{abstract}
Altermagnetism, a newly discovered magnetic phase, has spurred growing research activity. Studies from a perspective of dynamical electronic correlation still remain scarce. Employing density functional theory plus dynamical mean-field theory (DFT+DMFT) that incorporates dynamical electronic correlation, we demonstrate that CaCrO$_3$ is a strongly correlated altermagnet. Our DFT+DMFT calculations successfully reproduce the correlated metallic behavior of CaCrO$_3$ and quantitatively capture the incoherent state observed experimentally. We also identify that the altermagnetic CaCrO$_3$ is a Hund's metal. The incoherent state is attributed to Hund's coupling, which gives rise to a non-Fermi liquid behavior. Moreover, we find that altermagnetism can induce flat bands, and these incipient flat bands are further promoted by the strong renormalization from Hundness, which further drives a heavy-fermion behavior. Hence, we establish CaCrO$_3$ as a strongly correlated altermagnet and propose that Hund's metals provide an ideal platform for investigating the interplay between electronic correlation and altermagnetism. Our work will promote the study of strongly correlated altermagnetism physics.
\end{abstract}

\pacs{}

\maketitle

Recently, based on the spin-symmetry classification regarding collinear magnets, a new kind of magnetism beyond ferromagnetism (FM) and antiferromagnetism (AFM), namely altermagnetism, was proposed~\cite{PhysRevX.12.031042,PhysRevX.12.040501}. The sublattices with antiparallel spins in altermagnets cannot be connected by fractional translation and space-inversion operations, but can be connected by rotation operation~\cite{doi:10.1098/rspa.1966.0211,LITVIN1974538,Litvin:a14103,doi:10.1126/science.174.4013.985,PhysRevB.105.064430}, which causes spin splitting without spin-orbital coupling but maintains zero total magnetic moment. Thus, altermagnets share both advantages of FM spin splitting and AFM macroscopic zero net magnetic moment, which is very favorable to the designs and applications of spintronic devices~\cite{Jungwirth2016,RevModPhys.90.015005,Smejkal2018,Yan2024,Chen2024,PhysRevLett.133.106601,PhysRevB.107.L161109,PhysRevB.111.094411}. 

On the other hand, magnetism usually shows a close relationship with electronic correlation~\cite{Malrieu2014}, altermagnetism is not an exception. The coexistence of both wide and narrow bands that is caused by complex electronic correlations in magnets may be difficult to describe by using the conventional density functional theory (DFT). Especially for metallic magnets, the method of DFT+$U$ brings simple static collective corrections of energy for correlated orbitals, which ignores dynamical effects and may give inaccurate descriptions~\cite{wan2024highthroughputsearchmetallicaltermagnets}. As for experimental studies, the band renormalization that comes from correlation effect may bring challenges in detecting spin-splitting bands~\cite{PhysRevLett.132.036702,PhysRevLett.133.206401}. Hence, using approach of strong correlation that goes beyond the conventional DFT method could better describe and understand altermagnetism, which would provide guidance in enlarging the family of altermagnetic materials. Nevertheless, there are few studies regarding altermagnetism from a perspective of strong correlation.

The family of perovskite $AB$O$_3$~\cite{Marezioa07677} with a GdFeO$_3$-type structural distortion, a typical quantum material candidate involving multi-degrees of lattice, charge, orbital, and spin, is also a good platform for strongly correlated altermagnetism. Specifically, the transition-metal $B$ atoms are located at the centers of octahedrons of O atoms, which could generate rich strong correlation physics. Additionally, the existence of  two different tilted $B$O$_6$ octahedrons in $AB$O$_3$ may break space translation and space-inversion symmetry, which fits the requirement for forming altermagnetism.

\begin{figure}[b]
\centering
\includegraphics[width=8.6cm]{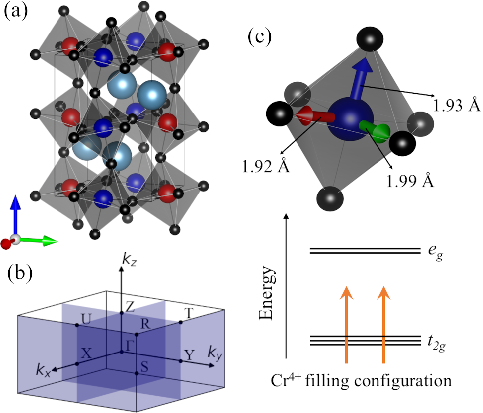}
\caption{(a) Crystal structure of CaCrO$_3$. The cyan, black, red, and blue balls denote atoms of Ca, O, spin-up Cr, and spin-down Cr, respectively. (b) Brillouin zone with the high-symmetry points of CaCrO$_3$. Spin-degenerate nodal planes are labeled in blue. (c) Schematic diagram regarding the filling configuration of the Cr-3$d$ orbitals in an octahedral crystal field.}
\label{fig:Fig1}
\end{figure}

\begin{figure*}[th]
\centering
\includegraphics[width=17.6cm]{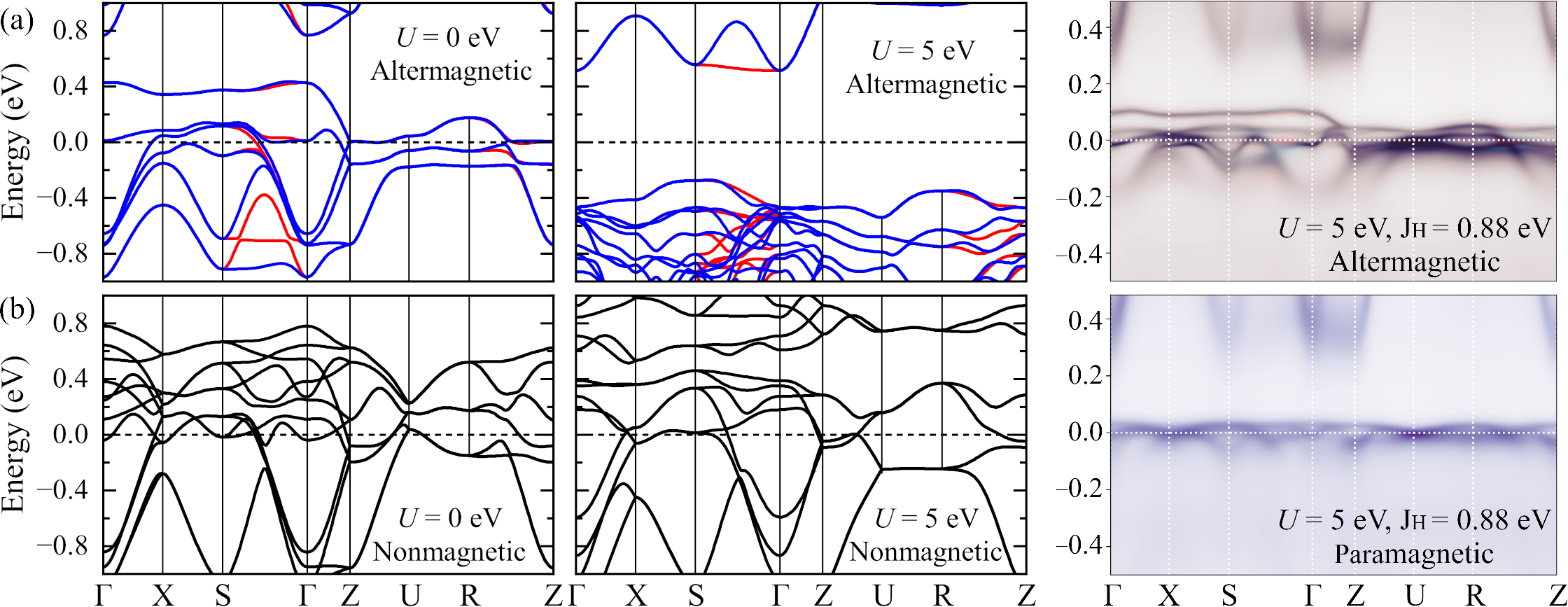}
\caption{Band structures of CaCrO$_3$ for (a) top panel: altermagnetic case and (b) bottom panel: nonmagnetic (paramagnetic) case. The red and blue spectrums in (a) denote the up and down spins. Left, middle, and right panels correspond DFT, DFT+$U$, and DFT+DMFT calculated results, respectively.   }
\label{fig:Fig2}
\end{figure*}

CaCrO$_3$ is a typical example of altermagnet~\cite{PhysRevLett.96.046408,PhysRevB.73.104409}, which had been confirmed to be a $C$-type AFM metal with a N\'{e}el temperature ($T_{\rm{N}}$) $\sim$ 90 K~\cite{PhysRevLett.101.167204,PhysRevB.83.165132}. After the concept of altermagnetism was proposed, a theoretical work based on DFT calculation and symmetry analysis studied anomalous Hall effect and spin splitting of CaCrO$_3$~\cite{PhysRevB.107.155126}. However, there is a non-negligible deviation between the DFT results and experimental results. Experimental results suggest that CaCrO$_3$ is a correlated antiferromagnetic metal~\cite{PhysRevB.83.165132}, while DFT+$U$ calculation shows a band insulating behavior with an FM ground state~\cite{PhysRevB.107.155126,PhysRevB.78.054425}. According to the studies on Hund's metal~\cite{PhysRevLett.107.256401,annurev}, CaCrO$_3$ could be a Hund's metal. Although DFT calculation could describe its altermagnetic metallic behavior, qualitatively~\cite{PhysRevB.107.155126},  ignoring electronic correlation effect is not a reasonable approximation. Given the number of known altermagnets is very limited~\cite{PhysRevX.14.031037,10.1093/nsr/nwaf066,adfm.202409327}, CaCrO$_3$ is a good platform for us to study altermagnetism from a perspective of strongly correlated physics.

In this Letter, we demonstrate that CaCrO$_3$ is a strongly correlated altermagnet as well as a Hund's metal with a non-Fermi liquid behavior. The correlated metallic behavior with altermagnetic spin-splitting of CaCrO$_3$ under the $T_{\rm{N}}$ is well reproduced within the framework of correction of dynamical mean-field theory (DMFT) and the incoherent state observed experimentally is quantitatively captured, for which the DFT+$U$ method fails. We also find that Hundness drives a heavy-fermion behavior by promoting the magnetism-induced incipient flat bands. Therefore, CaCrO$_3$ is a model material for investigating the interplay between Hundness and altermagnetic metal.

\textit{d-wave altermagnetism.}
As shown in Fig.~\ref{fig:Fig1}(a), CaCrO$_3$ has $Pbnm$ (No. 62) nonsymmorphic space group symmetry, with the corresponding point group generators being $C_{2x}$($\frac{1}{2}$, $\frac{1}{2}$, $\frac{1}{2}$), $C_{2y}$($\frac{1}{2}$, $\frac{1}{2}$, 0), and $I$. The Brillouin zone of CaCrO$_3$ is shown in Fig.~\ref{fig:Fig1}(b), where the high-symmetry points are labeled. Neutron scattering experiments indicate that the magnetic ground state of CaCrO$_3$ is $C$-type AFM, which is in-plane antiferromagnetic and out-of-plane ferromagnetic~\cite{PhysRevLett.101.167204,PhysRevB.83.165132}. Since the magnetic primitive cell and the crystal primitive cell are the same, CaCrO$_3$ breaks the \{$C_2$||$t$\} spin symmetry. On the other hand, Cr atoms with opposite spins are connected by $C_{2x}$($\frac{1}{2}$, $\frac{1}{2}$, $\frac{1}{2}$), $C_{2y}$($\frac{1}{2}$, $\frac{1}{2}$, 0), $M_{x}$($\frac{1}{2}$, $\frac{1}{2}$, $\frac{1}{2}$), and $M_{y}$($\frac{1}{2}$, $\frac{1}{2}$, 0), which endows CaCrO$_3$ with \{$C_2$||$C_{2x}$($\frac{1}{2}$, $\frac{1}{2}$, $\frac{1}{2}$)\},  \{$C_2$||$C_{2y}$($\frac{1}{2}$, $\frac{1}{2}$, 0)\}, \{$C_2$||$M_{x}$($\frac{1}{2}$, $\frac{1}{2}$, $\frac{1}{2}$)\}, and  \{$C_2$||$M_{y}$($\frac{1}{2}$, $\frac{1}{2}$, 0)\} spin symmetries. The breaking of the \{$C_2$||$I$\} and \{$C_2$||$t$\} spin symmetries can lead to spin splitting of bands. However, the spin symmetries \{$C_2$||$M_{x}$($\frac{1}{2}$, $\frac{1}{2}$, $\frac{1}{2}$)\} and \{$C_2$||$M_{y}$($\frac{1}{2}$, $\frac{1}{2}$, 0)\} can protect the spin degeneracy of the bands in the $k_x$ = 0 and $\pi$ planes, as well as the $k_y$ = 0 and $\pi$ planes as shown in Fig.~\ref{fig:Fig1}(c). Apart from these four planes, the bands of CaCrO$_3$ are spin-split. Hence, CaCrO$_3$ is a $d$-wave altermagnetic material.

\begin{table*}[thb]
\begin{center}
\small
\renewcommand\arraystretch{1.5}
\caption{DFT+DMFT calculated weights (\%) of the Cr-$t_{2g}$ orbitals local spin multiplets for altermagnetic (80 K) and paramagnetic (290 K) CaCrO$_3$, respectively.}
\label{tab1}
\begin{tabular*}{\linewidth}{@{\extracolsep{\fill}} ccccccccccc}
\hline\hline
{$N_{\Gamma}$}   & 1  & 1 & 2   & 2   & 2 & 3 & 3  & 3   &3   & other   \\
\hline
{$S_z$}  & 1/2  &$-$1/2    & 0   & 1  &$-$1   &1/2   &$-$1/2   &3/2   &$-$3/2   & other  \\
\hline
\makecell{Altermagnetism (80 K)}  & 3.87 &0.83 &5.29 & 47.29 &7.43 & 14.09 &2.97 & 13.20 &2.12 & 2.90  \\
\hline
\makecell{Paramagnetism (290 K)}  & 2.08 &2.08 &4.43 & 27.98 &27.98 & 8.65 &8.65 & 7.61 &7.61 & 2.94  \\
\hline\hline
\end{tabular*}
\end{center}
\end{table*}

\textit{Results of static electronic correlation.}
An experimental result of resonant photoemission spectroscopy (PES)~\cite{PhysRevB.83.165132} suggests that there is an on-site Hubbard $U$ $\sim$ 4.8 eV in CaCrO$_3$, which would be a correlated metallic system lying in the regime intermediate to Mott-Hubbard and charge-transfer systems. The DFT results in the left panels of Figs.~\ref{fig:Fig2}(a) and (b) show a metallic behavior of both altermagnetic and nonmagnetic CaCrO$_3$. Altermagnetism induces flat bands around the Fermi level. However, after considering static electronic correlation, the DFT+$U$ results of altermagneric CaCrO$_3$ shown in the middle panel of Fig.~\ref{fig:Fig2}(a) give an insulating behavior with a parameter of $U$ = 5 eV, which is close to the suggested value of the experiment. This calculated results are inconsistent with the experimental ones. So, we check the evolution behavior of the electronic structure with different values of $U$ (See in the SM~\cite{SM}). We find that the metallic behavior disappears and the insulating behavior appears when the $U$ is larger than 2 eV, which is consistent with a previous study~\cite{PhysRevB.107.155126}. More importantly, we find that the bandwidth of the low-energy band remains almost unchanged with the different values of $U$~\cite{SM}. The static electronic correlation just brings an overall shift to the bands around the Fermi level~\cite{SM}. Our calculated results indicate that the static electronic correlation method fails to describe altermagnetic CaCrO$_3$. An approach that goes beyond it is needed.

\textit{Results of dynamical electronic correlation.}
Here, we further study the electronic correlation of CaCrO$_3$ by performing DFT+DMFT calculations. We show the calculated spectral function $A(\bf{k}, \omega)$ of altermagnetic CaCrO$_3$ in the right panel of Fig.~\ref{fig:Fig2}(a). The correction brought by dynamical electronic correlation gives a scenario of a strongly correlated metal with significant band renormalization. The low-lying bands shrink into a narrow region that ranges from $\sim$ $-$0.2 to $\sim$ 0.1 eV with respect to the Fermi level. And the spin splitting that is determined by symmetry could still be captured along the lines of $S$-$\Gamma$ and $R$-$Z$, although the renormalized bandwidth is quite small (A spectral function $A_{up}({\bf{k}}, \omega) - A_{down}({\bf{k}}, \omega)$ is exhibited in the SM~\cite{SM}).

More importantly, the DFT+DMFT result differs from the DFT+$U$ result where the static electronic correlation causes an overall band shift and gives an insulating behavior. The DFT+DMFT calculated spectral function $A(\bf{k}, \omega)$ in the right panel of Fig.~\ref{fig:Fig2}(a) shows a correlated metallic behavior of altermagnetic CaCrO$_3$. Moreover, our findings are in quantitative agreement with experimental results. Specifically, the correlated electronic structure in the right panel of Fig.~\ref{fig:Fig2}(a) shows that there is incoherent electronic state at $\sim$ $-$0.2 eV below the Fermi level, which makes bands become too blurry to be visible. This finding is consistent with the PES result~\cite{PhysRevB.83.165132}, where an incoherence peak of photoemission spectrum is observed at the same energy region. The DFT+DMFT calculated magnetic moment of altermagnetic CaCrO$_3$ is $\sim$ 1.3 $\mu_B$/Cr, which is also in good agreement with the experimental value $\sim$ 1.2 $\mu_B$/Cr measured by neutron diffraction~\cite{PhysRevLett.101.167204}. In the right panel of Fig.~\ref{fig:Fig2}(b), we exhibit the correlated electronic structure of high-temperature (290 K) paramagnetic CaCrO$_3$, which also shows an incoherent metallic behavior with more strong band renormalization. All these results indicate that the DFT+DMFT method well describes the correlated electronic structure of altermagnetic CaCrO$_3$.

\textit{Altermagnetic Hund's metal.}
As a representative example of perovskite $AB$O$_3$ family, CaCrO$_3$ has a filling configuration of two electrons in three orbitals [Fig.~\ref{fig:Fig1}(c)]. The configuration of one electron or hole away from half-filling serves as a fertile ground for Hundness~\cite{PhysRevLett.107.256401,annurev}. Many siblings of CaCrO$_3$ in the $AB$O$_3$ family are proposed to be Hund's metals, such as SrRuO$_3$ and SrCrO$_3$. In Table~\ref{tab1}, we list the local spin multiplets of altermagnetic (80 K) and paramagnetic (290 K) CaCrO$_3$. The high-spin state $S_z$ = 1 wins the competition by a large margin, which strongly indicates that altermagnetic CaCrO$_3$ is a Hund's metal. Correlated metallic systems that are dominated by Hundness show a special spin screening process at a region of relatively high temperatures, where the systems usually show a non-Fermi liquid behavior with incoherent band structures~\cite{annurev,PhysRevLett.101.166405,Deng2019}. The incoherence features of Hund's metal are observed in the high-temperature spectral function $A(\bf{k},\omega)$ of CaCrO$_3$ in the right panel of Fig.~\ref{fig:Fig2}(b). Specifically, Hund's coupling favors a configuration with parallel spins in different orbitals within a site. With decreasing temperature, local spin moments contributed by parallel spins are being gradually screened by itinerant electrons. After the spins are fully screened below a certain temperature, the system transforms into a Fermi liquid, which is similar to the Kondo physics.

\begin{figure}[b]
\centering
\includegraphics[width=8.6cm]{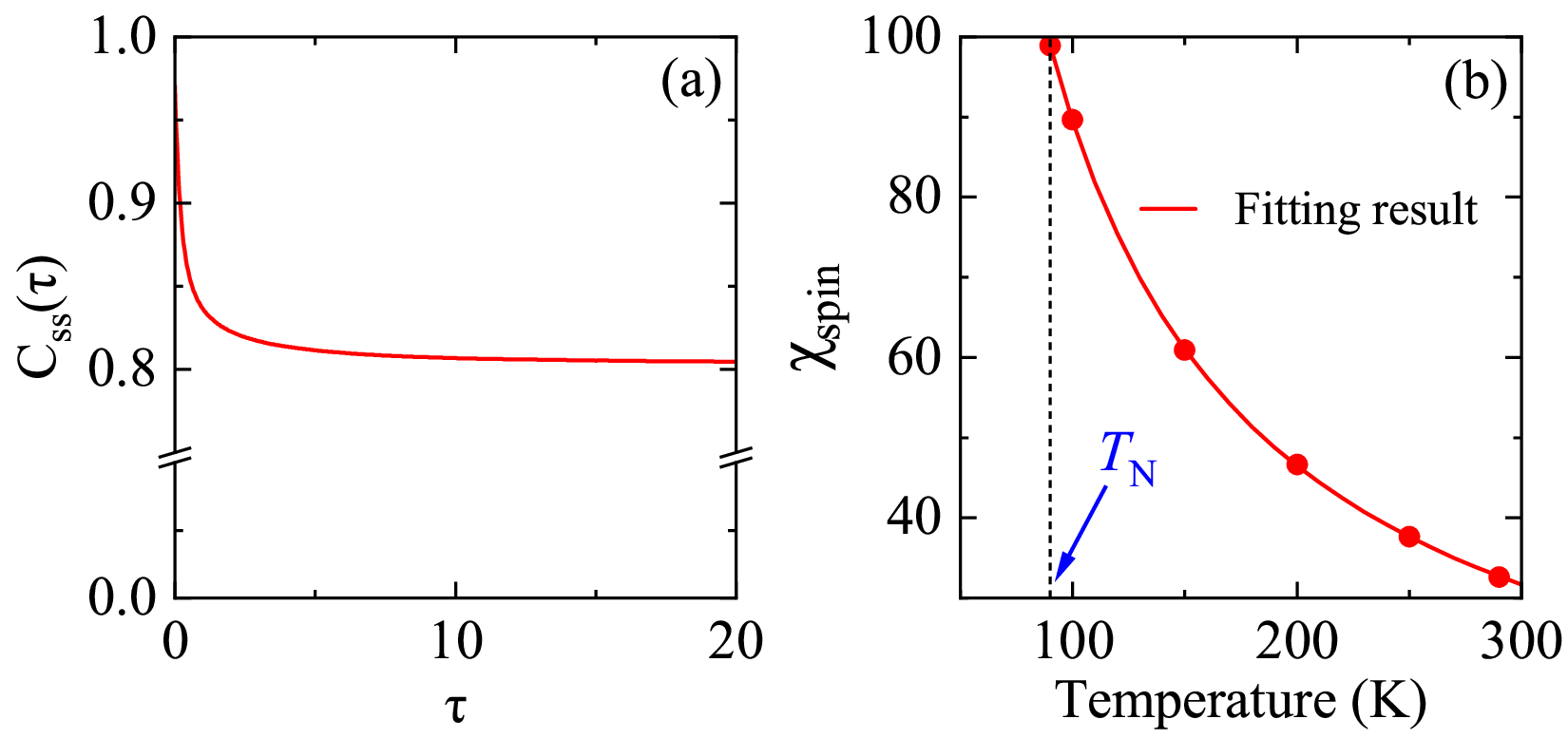}
\caption{(a) DFT+DMFT calculated imaginary-time spin-spin correlation function $C_{ss}(\tau)$ of paramagnetic CaCrO$_3$ at 290 K. (b) Static local spin susceptibilities $\chi_{\rm{spin}}$ of CaCrO$_3$ under different temperatures. The red line denotes the fitting result of $\chi$ $\sim$ 1/$T$.}
\label{fig:Fig3}
\end{figure}

\textit{Spin frozen phase.}
We show the imaginary-time spin-spin correlation function $C_{ss}(\tau)$ of CaCrO$_3$ at 290 K in Fig.~\ref{fig:Fig3}(a). The $C_{ss}(\tau)$ is defined as $C_{ss}(\tau)$ = ${\langle}S_z(\tau)S_z(0){\rangle}$, where $S_z$ is the $z$ component of spin. A large finite value of $C_{ss}(\tau)$ remains after an imaginary-time evolution of $\tau$ = $\beta$/2. This finding strongly indicates the existence of local spin moments in CaCrO$_3$ at a temperature higher than the $T_{\rm{N}}$ $\sim$ 90 K. The phase that involves unscreened spin degrees of freedom is known to be a spin frozen phase~\cite{annurev,PhysRevLett.101.166405}, where a non-Fermi liquid behavior and incoherent electronic structure usually appear. Similar findings have also been observed in other typical Hund's metals such as Sr$_2$RuO$_4$~\cite{Deng2019,PhysRevLett.106.096401} and recent Ruddlesden-Popper superconducting nickelates~\cite{PhysRevB.109.115114,PhysRevB.109.165140,PhysRevB.111.125111,1412-nfzm}. In Fig.~\ref{fig:Fig3}(b), the local spin susceptibility $\chi_{\rm{spin}}$ obeys the Curie-Weiss behavior $\chi \sim 1/T$ above the $T_{\rm{N}}$, which indicates the existence of local spin moments, a prerequisite for altermagnetism. Hence, we suggest that Hund's metals provide an ideal platform for hosting strongly correlated altermagnetism.

\begin{figure}[t]
\centering
\includegraphics[width=8.6cm]{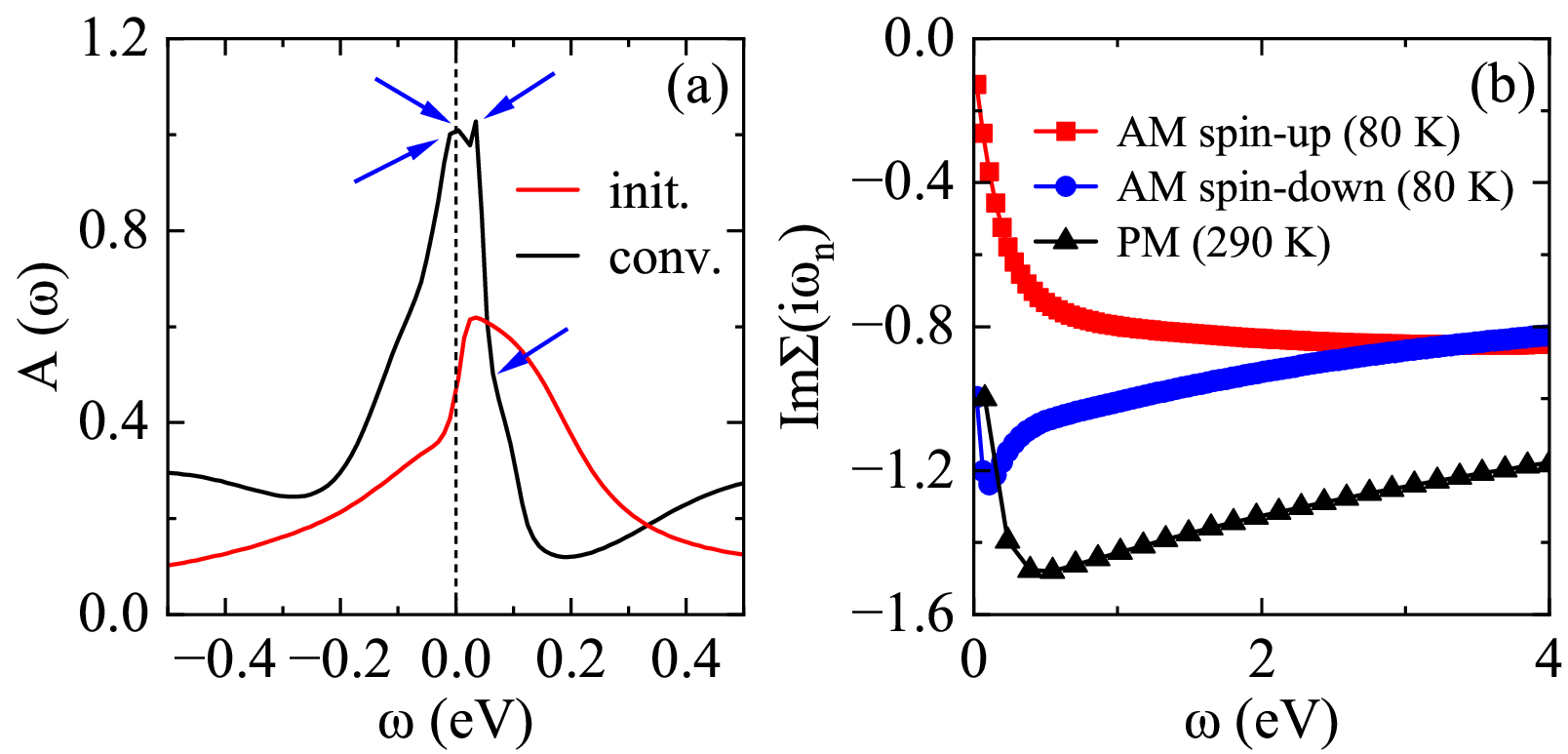}
\caption{(a) Initial and converged spectral functions $A(\omega)$ of the Cr-$t_{2g}$ orbitals of the self-consistent DFT+DMFT calculation. (b) Imaginary parts of the self-energy functions Im$\Sigma$($i\omega_n$) of the Cr-$t_{2g}$ orbitals of the altermagnetic (AM) and paramagnetic (PM) cases.}
\label{fig:Fig4}
\end{figure}

\textit{Promotion of flat bands.}
Here, we further investigate the interplay of electronic correlation and altermagnetism. The suppression of the coherence scale due to Hundness could promote incipient flat bands~\cite{PhysRevLett.106.096401}. In the DFT framework for CaCrO$_3$, the consideration of altermagnetism causes the emergence of flat bands around the Fermi level as shown in the left and middle panels of Figs.~\ref{fig:Fig2}(a) and (b). The correction brought by dynamical electronic correlation further significantly promotes these incipient flat bands, which is exhibited in the right panel of Fig.~\ref{fig:Fig2}(a). In Fig.~\ref{fig:Fig4}(a), we show the initial and converged spectral functions $A(\omega)$ of the DFT+DMFT self-consistency circle. At the very beginning of the self-consistency circle, the imaginary part of the self-energy function is zero, hence the initial spectral function $A(\omega)$ can be regarded as a non-interacting limit. The comparison between the initial and converged spectral functions $A(\omega)$ suggests that dynamical electronic correlation turns the wide and smooth low-energy density of states (DOS) in the non-interacting case into a strongly renormalized DOS. The marked peaks of the converged spectral function $A(\omega)$ in Fig.~\ref{fig:Fig4}(a) correspond to those flat bands around the Fermi level in the right panel of Fig.~\ref{fig:Fig2}(a).

This promotion of incipient flat bands caused by Hund's coupling is expected to induce effective heavy fermions in 3$d$-electron systems. In Fig.~\ref{fig:Fig4}(b), we show the imaginary parts of the Matsubara self-energy functions Im$\Sigma(i\omega_n)$ of the Cr-$t_{2g}$ orbitals of altermagnetic and paramagnetic cases, respectively. The nonlinear self-energy functions Im$\Sigma(i\omega_n)$ at low-frequency region not only indicates a non-Fermi liquid behavior that is usually accompanied by the emergence of spin frozen phase~\cite{annurev,PhysRevLett.101.166405}, but also suggests a large mass enhancement in CaCrO$_3$. Furthermore, by defining mass enhancement as $m^*/m$ $\approx$ 1$-\frac{{\rm{Im}}(i\omega_0)}{\omega_0}$, where $\omega_0$ = $\pi$/$\beta$~\cite{PhysRevB.99.045122}, we find that the maximum mass enhancement $m^*/m$ of altermagnetic CaCrO$_3$ is $\sim$ 47. It should be noted that the mass enhancement of CaCrO$_3$  is comparable with that of those usual heavy-fermion systems with 4$f$ electrons, such as Cerium~\cite{PhysRevB.99.045122}, within the framework of DFT+DMFT. And similar findings have also been reported in some kagome chromium compounds, such as CsCr$_3$Sb$_5$~\cite{PhysRevB.111.035127} and CsCr$_6$Sb$_6$~\cite{song2024realizationkagomekondolattice,PhysRevB.111.205120}. But the difference is that the incipient flat bands of CaCrO$_3$ here come from altermagnetism, in comparison, the incipient flat bands of CsCr$_3$Sb$_5$ and CsCr$_6$Sb$_6$ are caused by structural frustration of the kagome lattice.

\textit{Conclusion.}  
We demonstrate that CaCrO$_3$ is a strongly correlated altermagnet by performing DFT+DMFT calculations. Our results correctly describe the correlated altermagnetic metallic behavior and well reproduce incoherent state observed experimentally, while the DFT+$U$ method fails. Moreover, we find that CaCrO$_3$ is a Hund's metal. The experimentally observed incoherent state is attributed to Hund’s coupling, which drives a non-Fermi liquid behavior. The band renormalization brought by Hundness could promote the incipient flat bands induced by altermagnetism and further drive a heavy-fermion behavior in altermagnetic CaCrO$_3$. Therefore, we propose that Hund’s metals provide an ideal platform for investigating strongly correlated altermagnetism with novel physical properties.

Since the discovery of altermagnetism, it rapidly sparked many investigations that combine altermagnetism with other physics as well as potential applications~\cite{PhysRevLett.133.156702,PhysRevLett.134.166701,PhysRevLett.132.056701,PhysRevLett.133.106601,PhysRevLett.134.146001,PhysRevLett.133.086503,PhysRevLett.134.176902,PhysRevLett.133.146602,PhysRevLett.133.166701,PhysRevLett.132.263402,PhysRevLett.134.106802,PhysRevLett.132.236701,PhysRevLett.133.196701,PhysRevLett.133.206702,PhysRevLett.133.056401,PhysRevLett.134.106801,PhysRevLett.133.226002,PhysRevLett.133.106701,PhysRevLett.130.216702,PhysRevMaterials.8.L041402,PhysRevLett.134.086701,Ma2021,khodas2025tuningaltermagnetismstrain,daghofer2025altermagneticpolarons,delre2025diracpointstopologicalphases}. Studying altermagnetism from a perspective of strongly correlated material calculation is crucial and necessary. On the one hand, it no doubt will push for the discovery of more altermagnets. Specifically, the band renormalization due to strong correlation could further bring difficulties to the experimental characterization of small altermagnetic spin splitting. Hence, strong-correlation theoretical methods, such as DFT+DMFT, may be indispensable. On the other hand, it will also help developments of crossing fields involving altermagnetism. For example, spin splitting caused by altermagnetism indicates that altermagnet can realize spin Hall effect. Berry curvature, a key physical quantity to Hall effect, is generally calculated through DFT (DFT+$U$) method. However, our work here suggests that dynamical correlation is more suitable than the conventional static correlation to describes renormalized band structures of strongly correlated altermagnets. It is expected that this will inspire more experimental investiagtions regarding strongly correlated altermagnet, as well as corresponding theoretical developments such as strongly correlated topology, etc.

\begin{acknowledgments}
This work was supported by the National Key R\&D Program of China (Grants No. 2024YFA1408601 and No. 2024YFA1408602) and the National Natural Science Foundation of China (Grant No. 12434009). 
Z.O. was also supported by the Outstanding Innovative Talents Cultivation Funded Programs 2025 of Renmin University of China. 
P.J.G. was also supported by the National Natural Science Foundation of China (Grant No.12204533). 
Z.Y.L. was also supported by the Innovation Program for Quantum Science and Technology (Grant No. 2021ZD0302402). 
Computational resources were provided by the Physical Laboratory of High Performance Computing in Renmin University of China.
\end{acknowledgments}

\nocite{PhysRevB.54.11169}
\nocite{PhysRevB.50.17953}
\nocite{PhysRevLett.77.3865}
\nocite{Blaha-JCP152}
\nocite{Haule-PRB81}
\nocite{Haule-PRB75}
\nocite{Haule-PRL115}
\nocite{Jarrell-PR269}

\bibliography {CaCrO3}

\end{document}